\documentclass[10pt,prl,twocolumn,superscriptaddress]{revtex4}

\usepackage{graphicx}
\usepackage{amssymb}
\usepackage{times}
\usepackage{amsmath}

\begin{document}
    
\title{Tripartite Quantum State Sharing}

\author{Andrew M. Lance} \affiliation{Quantum Optics Group, Department
of Physics, Faculty of Science, Australian National University,
ACT 0200, Australia}

\author{Thomas Symul} \affiliation{Quantum Optics Group, Department of
Physics, Faculty of Science, Australian National University, ACT
0200, Australia}

\author{Warwick P. Bowen} \affiliation{Quantum Optics Group,
Department of Physics, Faculty of Science, Australian National
University, ACT 0200, Australia}

\author{Barry C. Sanders} \affiliation{Quantum Information Science
Group, Department of Physics and Astronomy, University of Calgary,
Alberta T2N 1N4, Canada}

\author{Ping Koy Lam} \affiliation{Quantum Optics Group, Department of
Physics, Faculty of Science, Australian National University, ACT
0200, Australia}

\begin{abstract}

We demonstrate a multipartite protocol to securely distribute and 
reconstruct a quantum state.
A secret quantum state is encoded into a tripartite entangled state 
and distributed to three players. 
By collaborating, any two of the three players can reconstruct the state, 
whilst individual players obtain nothing.
We characterize this $(2,3)$ threshold quantum state sharing scheme 
in terms of fidelity, signal transfer and reconstruction noise.
We demonstrate a fidelity averaged over all reconstruction permutations of 
$0.73 \pm 0.04$, a level achievable only using quantum resources. 

\end{abstract}


\date{\today} \maketitle

Secret sharing\cite{Sha79} is a powerful technique in computer science, which 
enables secure and robust communication in information networks,
such as the internet, telecommunication systems and distributed computers. 
The security of these networks can be enhanced using quantum 
resources to protect the information. Such schemes have been termed  
{\it quantum secret sharing}\cite{Hil99}.
Many applications in quantum information science, however, 
require the distribution of quantum states. 
One such example are quantum information networks, 
which are expected to consist of nodes where quantum states 
are created, processed and stored, connected by quantum 
channels\cite{Cir97}. 
It is of paramount importance that the 
quantum channels in these networks allow the robust and secure 
distribution of quantum states between nodes. 
Cleve {\it et al.}\cite{Cle99} proposed the secret sharing of 
quantum states as a protocol that provides these 
capabilities, overcoming failures or conspiracies by nodes. 
We term this {\it quantum state sharing} to differentiate from 
the quantum secret sharing of classical information. 
In $(k,n)$ threshold quantum state sharing\cite{Cle99}, the ``dealer'' node 
encodes a secret state into an $n$-party entangled state and
distributes it to $n$ ``player'' nodes. Any $k$ players (the access 
structure) can collaborate to retrieve the quantum state, whereas the 
remaining $n-k$ players (the adversary structure), even when 
conspiring, acquire nothing. This scheme provides quantum 
information networks with a secure framework for distributed quantum 
computation and quantum communication.

The original quantum state sharing scheme by Cleve {\it et al.} 
was formulated for discrete states and requires the control and coupling of qudits 
($d$-dimensional extensions of qubits), which is extremely experimentally challenging.      
In the continuous variable regime, however, quantum state sharing is 
feasible utilizing Einstein-Podolsky-Rosen (EPR) 
entanglement\cite{Tyc02}, an 
experimentally accessible quantum resource\cite{Ou92,Bow03}.
We demonstrate $(2,3)$ threshold quantum state 
sharing in this regime. In our scheme, a 
secret coherent state is encoded into a tripartite 
entangled state and distributed to three players. 
We demonstrate that any two of the three players can form an access structure to reconstruct the state. The state reconstruction is characterized in terms of fidelity, signal transfer, and reconstruction noise. 
These measures show a direct verification of our tripartite continuous variable entanglement. As coherent states form an over-complete basis for all quantum states, arbitrary states can be shared by this scheme.  

\begin{figure}[b]
\includegraphics[width=9cm]{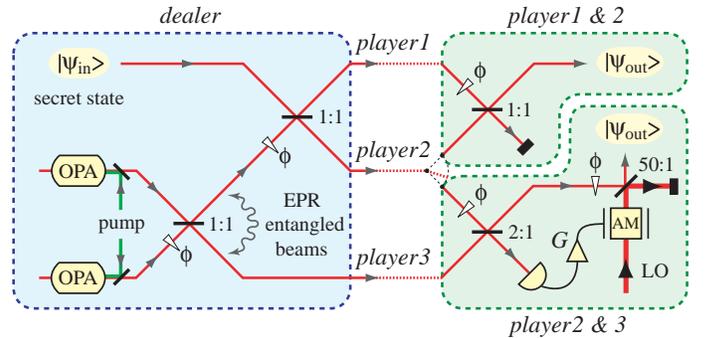}
\caption{ Schematic of the $(2,3)$ quantum state sharing
scheme.  $\psi_{\rm in}$: secret quantum state, OPA:
optical parametric amplifier, $G$: electronic gain, AM:
amplitude modulator, LO: optical local oscillator. ${\rm
x\!\!:\!\!y}$: beam splitter with reflectivity ${\rm x}/\!({\rm x}\!+\!{\rm y})$ and 
transmitivity ${\rm y}/\!({\rm x}\!+\!{\rm y})$.}\label{fig:QSSexperiment}
\end{figure}

The quantum states of interest in this paper reside at the frequency sidebands of an
electromagnetic field. In the Heisenberg picture of
quantum mechanics, a quantum state can be represented by the field
annihilation operator  $\hat{a}\!=\!(\hat X^{+}\!+\!i\hat
X^{-})/2$, where $\hat X^{\pm}\!=\!\langle\hat
X^{\pm}\rangle\!+\!\delta\hat X^{\pm}$ are the amplitude (+) and phase (-)
quadratures, with variances of $V^{\pm}\!=\!\langle(\hat X^{\pm})^{2}\rangle$.
In our dealer protocol, the dealer
interferes the secret state $\hat{a}_{{\rm in}}$ with one of a pair of
EPR entangled beams $\hat{a}_{\rm EPR1}$ on a 1:1 beam splitter
 (Fig.~\ref{fig:QSSexperiment}). The two output fields and the
second entangled beam $\hat{a}_{\rm EPR2}$ form the three shares which
are distributed to the players. The entangled state ensures that the
secret is protected from each player individually and is generated by
interfering a pair of amplitude squeezed
beams $\hat{a}_{\rm sqz1}$ and
$\hat{a}_{\rm sqz2}$\cite{Bow03}. The dealer can further enhance the
security of the scheme by displacing the coherent amplitudes of the
shares with correlated Gaussian white noise\cite{Lan03}.
Choosing the Gaussian noise to have the same correlations as the
quadrature entanglement, the shares can be expressed as
\begin{eqnarray}
\hat a_1 &=& (\hat a_{{\rm in}}\!+\!\hat a_{\rm EPR1}\!+\!\delta\mathcal{N}
)/\sqrt2 \\
\hat a_2 &=& (\hat a_{{\rm in}}\!-\!\hat a_{\rm EPR1}\!-\!\delta\mathcal{N}
)/\sqrt2 \\
\hat a_3 &=& \hat a_{\rm EPR2}\!+\!\delta\mathcal{N}^{\ast}
\end{eqnarray}
where $\delta\mathcal{N}\!=\!(\delta{N}^{+}\!+\!i\delta{N}^{-})/2$ represents the
Gaussian noise with mean $\langle\delta{N}^{\pm}\rangle\!=\!0$
and variance $\langle(\delta{N}^{\pm})^{2}\rangle\!=\!V_{N}$, and
$^\ast$ denotes the complex conjugate.

The reconstruction protocol used for the $(2,3)$ quantum state
sharing scheme is dependent on the corresponding access structure
(Fig.~\ref{fig:QSSexperiment}). The access structure
formed when players 1 and 2 collaborate,
henceforth denoted \{1,2\}, reconstructs the secret
quantum state by completing a Mach-Zehnder interferometer
using a 1:1 beam splitter\cite{Tyc02}. The access structures \{2,3\} and \{1,3\}
reconstruct the quantum state by utilizing a 2:1 beam splitter and
an electro-optic feedforward loop\cite{Lan03}. In the latter
protocol, combining the shares on the beam splitter with
appropriate relative phase reconstructs the phase quadrature of
the secret state on one of the beam splitter outputs.
In contrast, the amplitude quadrature has additional noise as a result of this
process. This noise, however, is correlated with the amplitude
quadrature of the other beam splitter output, which is detected.
The resulting photocurrent is fedforward to displace the amplitude
quadrature of the first output. Assuming no losses, the quadratures of
the reconstructed secret can then be expressed as \cite{Lan03}
\begin{eqnarray} \label{eqn:23reconstruction}
\delta\hat X^+_{{\rm out}} &=& \!g^{+}\delta\hat{X}^+_{{\rm
in}}\!+\!\frac{\sqrt{3}}{2}(1\!-\!\sqrt{3}g^{+})(\delta\hat{X}^+_{\rm
sqz_1}\!+\!\delta\hat{X}^+_{\rm sqz_2})+\nonumber\\
& & \!\!\frac{1}{2}(g^{+}\!\!-\!\sqrt{3})(\delta\hat X^-_{\rm
sqz_1}\!\!-\!\delta\hat X^-_{\rm
sqz_2})\!+\!(\!\sqrt{3}\!-\!g^{+}\!)\delta N^+\\
\delta\hat X^-_{{\rm out}} &=& \!\frac{1}{\sqrt{3}} (\delta\hat
X^-_{{\rm in}}\!+\!\delta\hat{X}^+_{\rm sqz_1}\!-\!\delta\hat
{X}^+_{\rm sqz_2})
\end{eqnarray}
where $g^{\pm}\!=\!\langle\hat X^{\pm}_{{\rm
out}}\rangle /\langle\hat X^{\pm}_{{\rm in}}\rangle$ are the
optical quadrature gains. The phase quadrature gain
$g^{-}\!=\!1/\sqrt{3}$ is set by the 2:1 beam splitter, whilst the amplitude
quadrature gain $g^{+}\!=\!(1/\sqrt{3}\!+\!G/\sqrt{6})$ has an additional term 
which is a function of the electronic feedforward gain $G$. We refer to
the specific gain of $g^{+}g^{-}\!=\!1$ as the \textit{unitary gain point}.
At unitary gain and in the limit of
perfect squeezing, the quadratures of the reconstructed state are
given by $\delta\hat{X}^{\pm}_{{\rm out}}\!=\! (\sqrt{3})^{\pm
1}\delta\hat{X}^{\pm}_{{\rm in}}$. This
state is directly related to the secret state via a local unitary
parametric operation. Although not in the same form as the secret state, such a reconstructed state is only achievable using entanglement. 
On the other hand, the unitary parametric operation is local and requires no entanglement. Therefore, the essence of the reconstruction protocol is contained within the feedforward scheme. 

\begin{figure}[b]
\includegraphics[width=8.5cm]{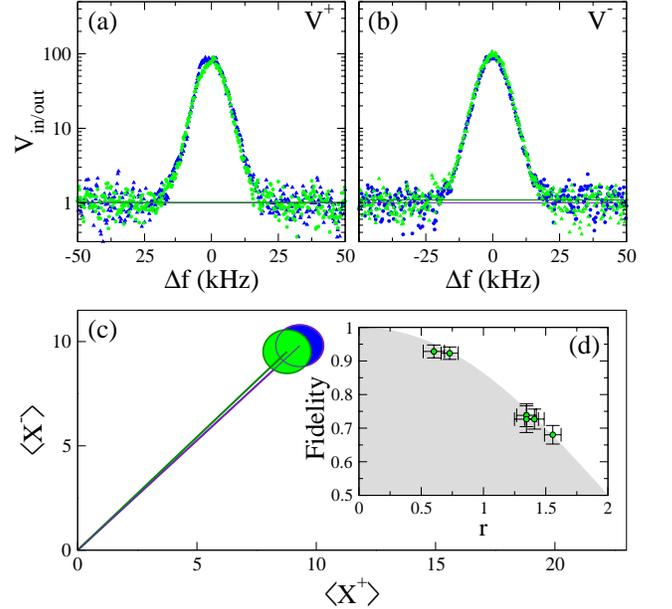}
\caption{Experimental results for the \{1,2\} access structure.
(a) and (b) show the spectra of the amplitude and phase quadrature
variances for the secret (input, blue/dark grey) and reconstructed (output, green/light grey)
quantum states. $\Delta{f}$ is the offset from the signal frequency at 6.12~MHz. 
Resolution Bandwidth = 1~kHz,
Video Bandwidth = 30~Hz. 
(c) Standard deviation contours of
Wigner functions of the secret (blue/dark grey) and reconstructed (green/light grey)
quantum states. (d) Measured fidelity as a function of gain
deviation $r^{2}\!=\!(\langle\!\hat{X}^{+}_{\rm out} \!  \rangle
\!  - \! \langle \!  \hat{X}^{+}_{\rm in} \!  \rangle)^{2} \!  +
\! (\langle \!  \hat{X}^{-}_{\rm out} \!  \rangle \!  - \! \langle
\! \hat{X}^{-}_{\rm in} \!  \rangle)^{2}$. (d) Grey area highlights the
accessible fidelity region.  Points plotted are from six different
experimental runs.} \label{fig:Fig_spectra3}
\end{figure}

\begin{figure}[b]
\includegraphics[width=8.5cm]{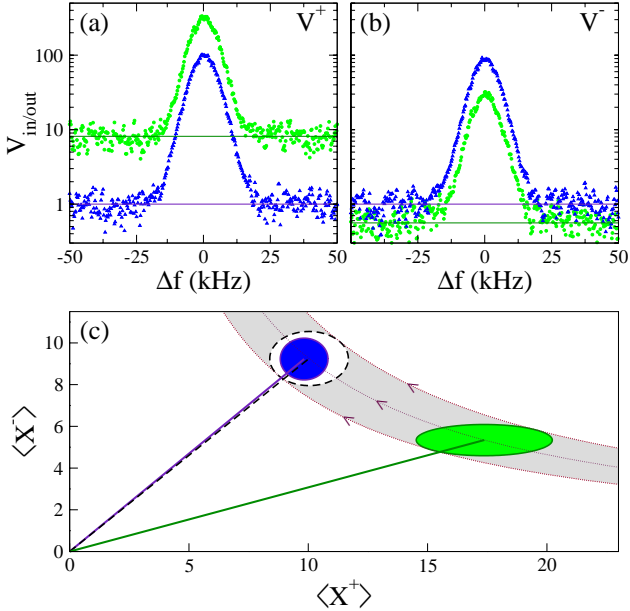}
\caption{Experimental results for the \{2,3\} access structure.
(a) and (b) show the spectra of the amplitude and phase quadrature
variances for the secret (input, blue/dark grey) and reconstructed 
(output, green/light grey) quantum states. 
(c) Standard deviation contours
of Wigner functions of the secret (blue/dark grey) and reconstructed (green/light grey)
quantum states. The dashed circle represents the quantum 
state $\delta\hat{X}^{\pm}_{\rm para} \!\!  =
\!\!  (\sqrt{3})^{\mp 1}\delta\hat{X}^{\pm}_{\rm out}$ 
after the a posteriori local unitary parametric operation.} \label{fig:Fig_spectra2}
\end{figure}

In our experiment we use a Nd:YAG laser producing a coherent field
at 1064nm to provide a shared time frame between all parties; to
yield the dealer secret state, produced by displacing the sideband
vacuum state of the laser field using an amplitude and a phase
modulator at 6.12MHz; and to produce two amplitude squeezed states
generated in hemilithic MgO:LiNbO$_{3}$ optical parametric
amplifiers (OPAs) pumped with 532nm light.
The output fields of each OPA are squeezed $4.5\pm{0.2}$dB below the
quantum noise limit. 
These squeezed beams are interfered on a 1:1 beam
splitter with an observed visibility of $99.1\pm{0.2}\%$. The beam splitter outputs are
EPR entangled and satisfy the wave-function
inseparability criterion
$\langle(\delta\hat{X}^{+}_{\rm EPR1}\!+\!\delta\hat{X}^{+}_{\rm
EPR2})^{2}\rangle\langle(\delta\hat{X}^{-}_{\rm
EPR1}\!-\!\delta\hat{X}^{-}_{\rm
EPR2})^{2}\rangle/4\!=\!0.44\pm{0.02}\!<\!1$\cite{Dua00,Bow03}.
To enhance the security of the secret state against the adversaries, the coherent quadrature
amplitudes of the entangled beams are displaced with Gaussian
noise of variance $V_{N}=3.5\pm{0.1}$dB. 
Experimentally, this noise can be actively applied using electro-optic modulation techniques, but in our case it appears naturally as a result of de-coherence\footnote{This aspect of de-coherence will be described in detail in a later publication.}.
A homodyne detector is used
to characterize the secret, adversary and reconstructed quantum
states using a configuration of removable mirrors.
To ensure accurate results, the total
homodyne detection efficiency, $\eta_{\rm
hom}\!=\!0.89\!\pm\!{0.01}$, is inferred out of each measurement.

\begin{figure}[b]
\includegraphics[width=8.5cm]{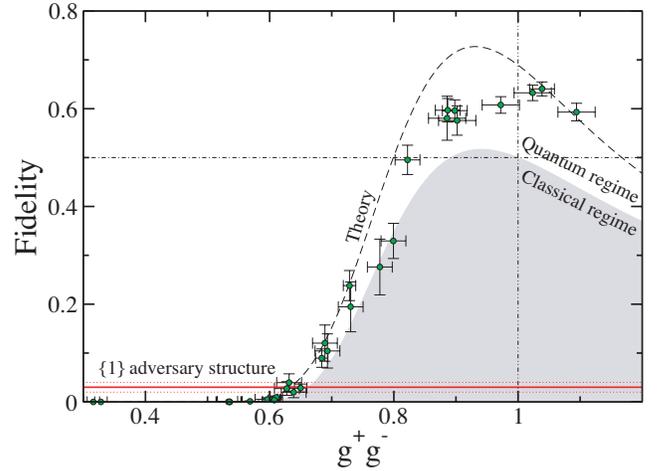}
\caption{ Experimental fidelity for the \{2,3\} access structure as
a function of the product of $g^{+}g^{-}$. Dashed line: calculated theoretical curve with
squeezing of $-4.5$ dB, added noise of $+3.5$ dB, electronic
noise of $-13$ dB with respect to the quantum noise limit, and
feed forward detector efficiency of $0.93$.
Solid line and dotted lines: experimental
fidelity for the adversary structure and error bar. Grey area 
highlights the classical boundary for the access structure.} \label{fig:QSS_Fidelity}
\end{figure}

\begin{figure}[b]
\includegraphics[width=8.5cm]{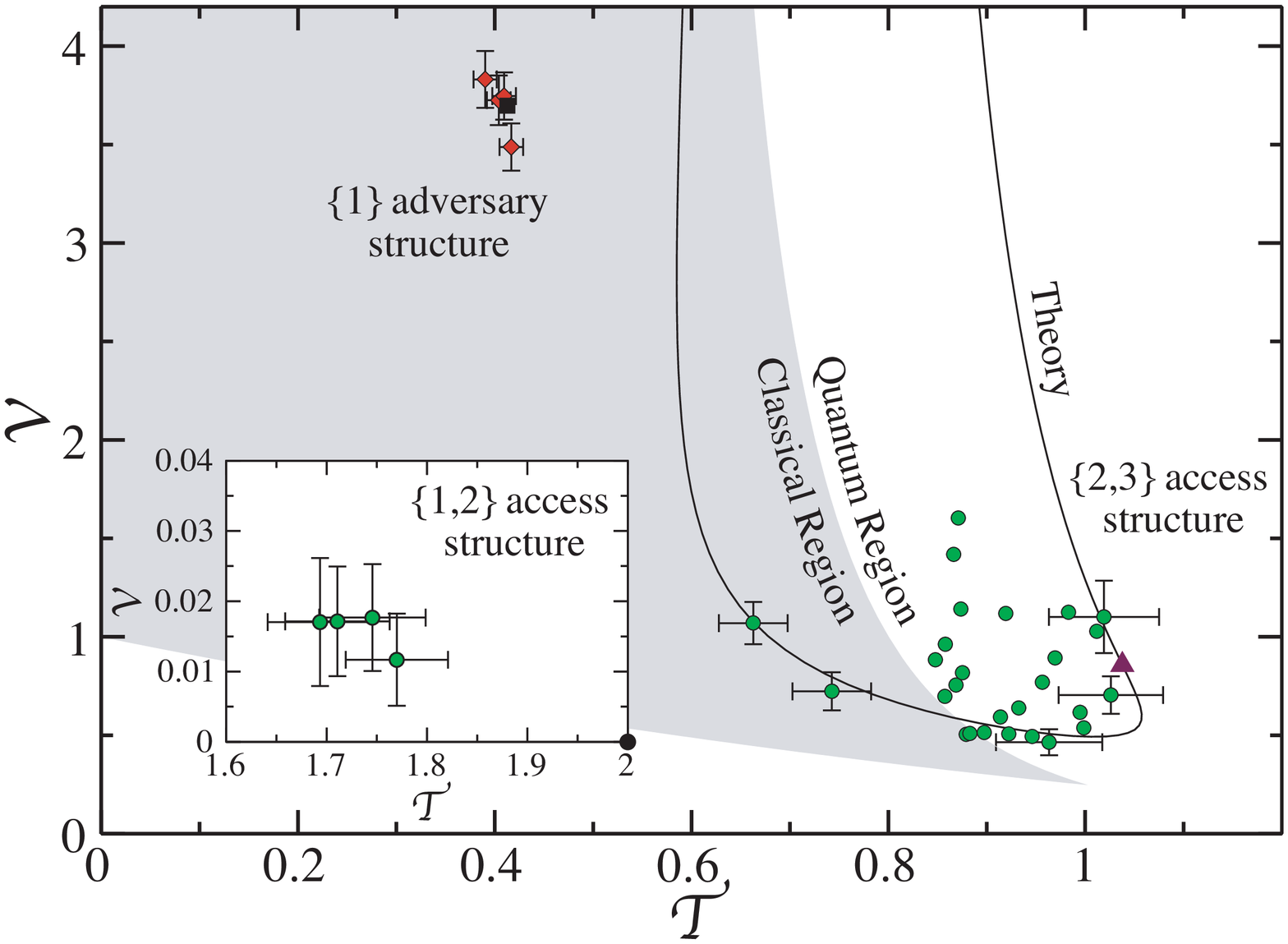}
\caption{Experimental signal transfer ($\mathcal{T}$) and additional 
noise ($\mathcal{V}$) for the \{2,3\} access 
structure (green/light grey circles), and the adversary structure (red/dark grey diamonds). Solid line: calculated theoretical curve for varying gain with same parameters as in Figure~\ref{fig:QSS_Fidelity}.
Triangle symbol: unitary gain point for the \{2,3\} access structure. 
Square symbol: calculated theoretical point for the adversary structure.
Grey area: the classical region for the \{2,3\} access structure.
(inset) Experimental $\mathcal{T}$ and $\mathcal{V}$
for the \{1,2\} access structure (green/light grey circles)
and the theoretical point (black circles).
} \label{fig:QSS_TV}
\end{figure}

We characterize the quality of the state reconstruction for the
access and adversary structures using fidelity
$\mathcal{F}\!=\!\langle \psi_{\rm in} |\rho_{\rm out}|\psi_{\rm
in}\rangle$, which measures the overlap between the secret and
reconstructed quantum states\cite{Shu95}. Assuming that all fields involved have
Gaussian statistics and that the secret is a coherent state, the
fidelity can be expressed in terms of experimentally measurable
parameters as
\begin{equation}
\mathcal{F}=2e^{-(k^{+}\!+\!k^{-})/4}\Big/ \sqrt{(1\!+\!V_{{\rm
out}}^{+})(1\!+\!V_{{\rm out}}^{-})}
\end{equation}
where $k^{\pm}\!=\!\langle X_{{\rm
in}}^{\pm}\rangle^{2}(1\!-\!g^{\pm})^{2}/ (1\!+\!V_{{\rm
out}}^{\pm})$. In our experiment, the fidelity for \{1,2\} can be determined
directly; however, for \{2,3\} and \{1,3\} 
a unitary parametric transformation must be applied before a meaningful fidelity is obtained. This unitary transform can be applied either optically or a posteriori. The final state is then 
$\delta\hat{X}^{\pm}_{{\rm para}}\!=\!  (\sqrt{3})^{\mp
1}\delta\hat{X}^{\pm}_{{\rm out}}$, so in the ideal case
$\delta\hat{X}^{\pm}_{{\rm para}}\!=\!\delta\hat{X}^{\pm}_{{\rm
in}}$. Under ideal conditions and at unitary gain, any one of the access
structures can achieve $\mathcal{F}\!=\!1$
corresponding to perfect reconstruction of the
secret quantum state, whilst the
corresponding adversary structure obtains $\mathcal{F}\!=\!0$.

When no entanglement is used, the 
maximum fidelity achievable by \{2,3\} and \{1,3\} is
$\mathcal{F}^{\rm clas}_{\{2,3\}}\!=\!\mathcal{F}^{\rm
clas}_{\{1,3\}}\!=\!1/2$, whilst \{1,2\} can still achieve
$\mathcal{F}^{\rm clas}_{\{1,2\}}\!=\!1$, so the average fidelity achieved by all
permutations of the access structure cannot exceed $\mathcal{F}^{\rm
clas}_{\rm avg}\!=\!2/3$. This defines the classical boundary for
$(2,3)$ threshold quantum state sharing.
Note that for general
$(k,n)$ threshold quantum state sharing of a coherent state, independent of the
dealer protocol, the maximum average fidelity achievable
without entanglement resources is $\mathcal{F}^{\rm clas}_{\rm
avg}\!=\!k/n$.

With the $\{1,3\}$ and $\{2,3\}$
protocols being equivalent, our $(2,3)$ threshold quantum state sharing scheme is demonstrated
through the implementations of the $\{1,2\}$ and $\{2,3\}$
reconstruction protocols.  Figure~\ref{fig:Fig_spectra3} shows
the noise spectra and corresponding inferred Wigner function standard deviation contours for
the secret and reconstructed state for the \{1,2\}
protocol.  The fidelity obtained from these noise spectra is
$\mathcal{F}_{\{1,2\}}\!=\!0.93\!\pm\!{0.03}$ with $g^{+}\!=\!0.94\!\pm\!{0.01}$ and
$g^{-}\!=\!0.97\!\pm\!{0.01}$.
The corresponding adversary structure \{3\} gets a fidelity of $\mathcal{F}_{\{3\}}\!=\!0$
since the share contains no component of the secret state.
Figure~\ref{fig:Fig_spectra3} (d) shows several measured fidelity
points as a function of phase space distance, $r$, between the coherent
amplitudes of the secret and reconstructed states. Each
fidelity point has a non-zero distance due to mode mismatch,
optical losses and imperfect phase locking.

Similarly,
Figure~\ref{fig:Fig_spectra2} shows an example of the secret and
reconstructed state for the \{2,3\} protocol. In this case, to allow a
direct comparison between the secret and reconstructed states, the inferred Wigner function standard deviation contour of
the reconstructed state after the a posteriori local unitary
parametric operation is also shown. Figure~\ref{fig:QSS_Fidelity}
shows the measured fidelity for a range gains.
Around the unitary gain point, we achieve a fidelity of
$\mathcal{F}_{\{2,3\}}\!=\!0.63\!\pm\!{0.02}$ with
$g^{+}g^{-}\!=\!1.02\pm{0.03}$.
The corresponding adversary structure \{1\} achieves a
fidelity of only $\mathcal{F}_{\{1\}}\!=\!0.03\!\pm\!{0.01}$.
The quantum nature of our protocol is demonstrated by the fidelity
averaged over all the access structures
$\mathcal{F}_{\rm avg}\!=\!0.74\!\pm\!{0.04}$, which exceeds the
classical limit $\mathcal{F}^{\rm clas}_{\rm avg}=2/3$.

 Fidelity is a single state dependent measure of the efficacy of 
quantum information protocols. There are alternative measures which 
provide complementary information about these processes. 
One obvious technique is to measure the signal transfer to 
($\mathcal{T}$), and the additional noise on ($\mathcal{V}$), 
the reconstructed state\cite{Ral98}.  
Such analysis has been used to characterize quantum 
non-demolition\cite{Poi94} and quantum 
teleportation experiments\cite{Bow03b}. 
Unlike the fidelity measure, both $\mathcal{T}$ and $\mathcal{V}$ 
are invariant to unitary transformations of the
reconstructed state. Therefore, for the $\mathcal{T}$ and $\mathcal{V}$ analysis, 
an a posteriori unitary transform is not required. 
The signal transfer is given by $\mathcal{T}\!=\!T^{+}\!+\!T^{-}$, where $T^{\pm}\!=\!{\rm SNR}^{\pm}_{\rm out}/{\rm SNR}^{\pm}_{\rm in}$ are the quadrature signal transfer coefficients, with ${\rm SNR}^{\pm}$ being the standard signal-to-noise ratios.
The additional noise is given by $\mathcal{V}\!=\!V^{+}_{\rm cv}V^{-}_{\rm cv}$, where 
$V^{\pm}_{\rm cv}\!=\!V^{\pm}_{\rm out}\!-\!|\langle\delta\hat{X}^{+}_{\rm in}
\delta\hat{X}^{+}_{\rm out}\rangle |^{2}/V^{\pm}_{\rm out}$ are the 
conditional variances. 
Experimentally, the signal-to-noise ratios that define $\mathcal{T}$ can be obtained from our measured noise spectra (Fig.~\ref{fig:Fig_spectra3} and \ref{fig:Fig_spectra2}), whilst 
$V^{\pm}_{\rm cv}$ can be determined from the output quadrature variance and the optical quadrature gains $V^{\pm}_{\rm cv}\!=\!V^{\pm}_{\rm out}\!-\!(g^{\pm})^{2}$.
In the ideal case, any one of the access structures can achieve 
perfect state reconstruction with $\mathcal{T}\!=\!2$ and 
$\mathcal{V}\!=\!0$, whilst the corresponding adversary structure 
obtains  no information with $\mathcal{T}\!=\!0$ and $\mathcal{V}\!=\!\infty$. 

Figure~\ref{fig:QSS_TV} (inset) shows the experimental $\mathcal{T}$ 
and $\mathcal{V}$ points for the \{1,2\} protocol. We 
measure a best state reconstruction of 
$\mathcal{T}_{\{1,2\}}\!=\!1.77\pm{0.05}$ and 
$\mathcal{V}_{\{1,2\}}\!=\!0.01\pm{0.01}$.   
Both of these values are close to optimal, being degraded only by 
optical losses and experimental inefficiencies.
Similarly, Figure~\ref{fig:QSS_TV} shows the points  
for the \{2,3\} protocol for a range of gains, together with the 
corresponding adversary structure \{1\}. 
The majority of the experimental points are in agreement with the theoretical 
prediction, with the discrepancies accountable for by drifts in our control system. 
The accessible region 
for the \{2,3\} protocol without entanglement is illustrated by the shaded region. 
The quantum nature of the state reconstruction is demonstrated by the 
experimental points which exceed this classical region. 
For the \{2,3\} protocol we measure a lowest
reconstruction noise of $\mathcal{V}_{\{2,3\}}\!=\!0.46\pm{0.08}$ 
and a largest signal transfer of $\mathcal{T}_{\{2,3\}}\!=\!1.03\pm{0.05}$. 
Points with $\mathcal{T}\!>\!1$ exceed the 
information cloning limit\cite{Bow03b} and demonstrate that the \{2,3\} protocol has 
better access to information encoded on the secret state than any other parties. 
The adversary structure obtains 
a mean $\mathcal{T}_{\{1\}}\!=\!0.41\pm{0.01}$ 
and $\mathcal{V}_{\{1\}}\!=\!3.70\pm{0.08}$. 
The separation of the adversary structure $\mathcal{T}$ and $\mathcal{V}$ points
from that of the \{2,3\} protocol in Figure~\ref{fig:QSS_TV} illustrates that 
in such a protocol the access structure performs far better than any 
adversary structure. 
   
Our experimental demonstration of $(2,3)$ threshold quantum state
sharing is the first application of continuous variable tripartite
entanglement.  
Furthermore, it is extendable to an arbitrary $(k,n)$ scheme, 
without a corresponding scale-up
of the required quantum resources\cite{Tyc03}. 
The implementation of quantum state sharing broadens the scope of quantum
information networks allowing quantum communication between multiple
nodes, whilst providing security against malicious parties in the
network as well as node and channel failures.  

We wish to thank Timothy Ralph, Tom\'a\v{s} Tyc, Roman Schnabel and Hans Bachor for useful discussions and the support of the Australian Research Council and iCORE.


\begin{thebibliography}{99}

\bibitem{Sha79} A.~Shamir, {\it Comm. of the ACM} {\bf 22}, 612 (1979).

\bibitem{Hil99} M.~Hillery, V.~Buzek and A.~Berthiaume,  {\it \pra} {\bf 59}, 1829 (1999);
 A.~Karlsson, M.~Koashi and N.~Imoto,  {\it \pra} {\bf 59}, 162 (1999);
W.~Tittel, H.~Zbinden and N.~Gisin,  {\it \pra} {\bf 63}, 042301 (2001).

\bibitem{Cir97} J.~I.~Cirac, P.~ Zoller, H.~J.~Kimble and H.~Mabuchi,  {\it \prl} {\bf 78}, 3221 (1997).

\bibitem{Cle99} R.~Cleve,  D.~Gottesman and H-K.~Lo, {\it \prl} {\bf 83}, 648 (1999).
 
\bibitem{Tyc02} T.~Tyc and B.~C.~Sanders, {\it \pra} {\bf 65}, 42310 (2002).

\bibitem{Ou92}  Z.~Y.~Ou, S.~F.~Pereira, H.~J.~Kimble and  K.~C.~Peng, {\it \prl}  {\bf 68}, 3663 (1992).

\bibitem{Bow03} W.~P.~Bowen, R.~Schnabel, P.~K~Lam  and T.~C.~Ralph, {\it \prl} {\bf 90}, 043601 (2003).

\bibitem{Lan03}  A.~M.~Lance, T.~Symul, W.~P.~Bowen, T.~Tyc, 
B.~C.~Sanders and P.~K.~Lam, {\it New J. of Phys.} {\bf 5}, 4.1 (2003).

\bibitem{Fur98} A.~Furusawa,  {\it et al.}, {\it Nature} {\bf 282}, 706 (1998). 

\bibitem{Dua00} L-M.~Duan, G.~Giedke, J.~I.~Cirac and P.~Zoller, 
 {\it \prl} {\bf 84}, 2722 (2000).

\bibitem{Shu95} B.~Schumacher,  {\it \pra} {\bf 51}, 2738 (1995).

\bibitem{Ral98} T.~C.~Ralph and P.~K.~Lam, {\it \prl} {\bf 81}, 5668 (1998)

\bibitem{Poi94} J.~-Ph.~Poizat, J.~-F.~Roch, and P.~Grangier, {\it Ann. Phys. (Paris)} {\bf 19}, 265 (1994).

\bibitem{Bow03b} W.~P.~Bowen, {\it et al.}, {\it \pra} {\bf 67},  032302 (2003). 

\bibitem{Tyc03} T.~Tyc, D.~J.~Rowe and  B.~C.~Sanders,  {\it J. Phys. A:Math. Gen.} {\bf36}, 7625 (2003).

\end{thebibliography}
\end{document}